\documentclass[usenatbib]{mnras}
\usepackage{latexsym}

\usepackage[dvipsnames]{xcolor}
\usepackage[title]{appendix}
\usepackage{graphicx}
\usepackage[space]{grffile}
\usepackage{latexsym}
\usepackage{longtable}
\usepackage{xspace}
\usepackage{amsfonts,amsmath,amssymb}
\usepackage{url}
\usepackage{hyperref}
\usepackage[utf8]{inputenc}
\usepackage[english]{babel}
\makeatletter
\newcommand*{\rom}[1]{\expandafter\@slowromancap\romannumeral #1@}
\newcommand{\hi}{{\sc H\,i}\xspace} 
\usepackage{color}
\usepackage{soul} 
\definecolor{red}{rgb}{1.0,0.2,0.2}
\definecolor{green}{rgb}{0.2,1.0,0.2}

\makeatother

\title[Towards the first detection of strongly lensed HI emission]
{Towards the first detection of strongly lensed HI emission}
\author[Blecher et al.]{Tariq Blecher$^{1,2}$, Roger Deane$^{3,1}$, Ian Heywood$^{4,1}$, Danail Obreschkow$^5$\\
~\\
\textit{$^1$ Centre for Radio Astronomy Techniques and Technologies, Department of Physics and Electronics, Rhodes University,-}\\
 \textit{Grahamstown 6140, South Africa}\\
\textit{$^2$  South African Radio Astronomy Observatory, Observatory 7925, Cape Town, South Africa}\\
\textit{$^3$ Department of Physics, University of Pretoria, Hatfield, Pretoria, 0028, South Africa }\\
\textit{$^4$ Astrophysics, University of Oxford, Denys Wilkinson Building, Keble Road, Oxford, OX1 3RH, UK} \\
\textit{$^5$ International Centre for Radio Astronomy Research (ICRAR), M468, University of Western Australia, WA 6009, Australia}\\
}

\begin{document}

\maketitle
\begin{abstract}
We report interferometric observations tuned to the redshifted neutral hydrogen (\hi) 21cm emission line in three strongly lensed galaxies at $z \sim 0.4$ with the Giant Metrewave Radio Telescope (GMRT). One galaxy spectrum (J1106+5228 at z=0.407) shows evidence of a marginal detection with an integrated signal-to-noise ratio of 3.8, which, if confirmed by follow-up observations, would represent the first strongly lensed and most distant individual galaxy detected in \hi emission. Two steps are performed to transcribe the lensed integrated flux measurements into \hi mass measurements for all three target galaxies. First, we calculate the \hi magnification factor $\mu$ by applying general relativistic ray-tracing to a physical model of the source-lens system. The \hi magnification generally differs from the optical magnification and depends largely on the intrinsic \hi mass $M_{\rm HI}$ due to the \hi mass-size relation. Second, we employ a Bayesian formalism to convert the integrated flux, amplified by the $M_{\rm HI}$-dependent magnification factor $\mu$, into a probability density for $M_{\rm HI}$, accounting for the asymmetric uncertainty due to the declining HI mass function (Eddington bias). In this way, we determine a value of $\log_{\rm 10} (M_{\rm HI}/M_\odot) = 10.2^{+0.3}_{-0.7}$ for J1106+5228, consistent with the estimate of $9.4\pm0.3$ from the optical properties of this galaxy. The \hi mass of the other two sources are consistent with zero within a 95 per cent confidence interval however we still provide upper limits for both sources and a $1\sigma$ lower limit for J1250-0135 using the same formalism.

\end{abstract}
\clearpage
\section{INTRODUCTION} 
Neutral atomic hydrogen (\hi) plays a key role in the baryon cycle of galaxies. Its spatial distribution within galaxies is diffuse and extended, with significant mass beyond the stellar component of the galaxy \citep{leroy_2008}. As the simplest, most abundant and spatially-extended galactic gas component, studies of neutral hydrogen can probe a wide range of astrophysics including star formation histories, galaxy interactions and cosmic large-scale structure. For example, the ratio of the cosmic densities of molecular to neutral hydrogen ${\rm \Omega_{H_2}/\Omega_{\rm HI}}$ is predicted to increase as a function of redshift, for instance as ${\rm \Omega_{H_2}/\Omega_{\rm HI}} \propto (1+z)^{1.6}$ between $0\le z\le 2$ \citet{obreschkow_2009b}. This results from the growth of haloes with cosmic time which leads to larger but less dense galactic discs. The decreasing density of galactic discs at lower redshift are then less efficient at converting ${\rm HI} \to {\rm H_2}$ due to the reduction in gas pressure. This decline in ${\rm \Omega_{H_2}/\Omega_{\rm HI}}$ with cosmic time parallels the rapid decrease in the co-moving star formation (SFR) rate density from $z\sim 2$ to the current epoch \citep[][and references therein]{hopkins_2006}.

Neutral hydrogen can be observed via the 21cm radio line, which results from the forbidden hyperfine (spin-flip) transition (we refer to this as the \hi line). Unfortunately, as the \hi emission line is extremely faint, it is difficult to constrain ${\rm \Omega_{\rm HI}}$ at $z>0.2$ with direct observations. Currently, the COSMOS \hi Large Extra-galactic Survey (CHILES) survey carried out on the Karl G. Jansky Very Large Array (VLA) holds the record for the most distant detection of \hi in emission from a single galaxy with $M_{\rm HI} = 2.9 \times 10^{10}$~M$_\odot$ at $z=0.37$ \citep{fernandez_2016}. The detection was reported using the first 178 hours of data of the 1002 hour survey of the COSMOS field. The redshift cutoff for the survey is at $z\sim 0.45$, governed by receiver sensitivity drop at the lower end of the VLA L-band, where L-band refers to the frequency band $\sim 1-2$~GHz. With the 305~m Arecibo dish, the HIghz  project \citep{catinella_2015} detected 39 galaxies $ 2 \le M_{\rm HI}/10^{10}\,{\rm M_\odot} \le 8$ at $0.17 \le z \le 0.25$. These \hi masses fall at the high end of the \hi mass function (HIMF) in the local Universe, well above the point at which the HIMF transitions into an exponential decline \citep{jones_2018}.

Absorption studies of the 21cm line \citep[e.g.][]{gupta_2013,allison_2015} and Lyman-$\alpha$ \citep[e.g.][]{prochaska_2011}, as well as statistical analyses such as stacking \citep{verheijen_2007,kanekar_2016} and intensity mapping \citep{chang_2010,masui_2013}, provide important constraints on high redshift \hi, but are strongly model-dependent techniques. Hence, direct detections are critical to cross-check and more directly constrain the high-redshift \hi mass function. Future radio telescopes like the Square Kilometre Array (SKA) and its pathfinders/precursors should be able to make detections of individual, massive galaxies towards $z\sim 1$. A promising route to  higher redshift \hi detections with current and future telescopes is the natural flux magnification enabled by strong gravitational lensing.

Amplification of emission through gravitational lensing has been used at many wavelengths to boost the signal of distant, faint galaxies, however there has been no strongly-lensed \hi detection in emission to date, with only a single published attempt known to the authors \citep{hunt_2016}. Predictions show that targeted observing campaigns should be able to detect lensed \hi within reasonable observing times of order a few days with current instruments \citep{deane_2015}. This would allow us to probe lower \hi mass galaxies at intermediate redshift, yielding complementary results to large-scale \hi surveys in progress/preparation.

In this paper, we present: (1) the results of Giant Metrewave Radio Telescope (GMRT) L-band observations of three galaxy-galaxy gravitational lenses, (2) Monte Carlo simulations of the ray-traced parameteric \hi discs to estimate the average (i.e. total \hi intensity) magnification and (3) a Bayesian formalism for a robust description of the \hi mass probability. In section~\ref{sec:obs}, we present the target selection, the observational details and data reduction; in section~\ref{sec:results} we present the interferometric data products (subsection~\ref{subsec:obs_result}), the \hi lensing simulations (subsection~\ref{subsec:sims}) and the \hi mass probability distributions (subsection~\ref{subsec:upper_limits}). In section~\ref{discussion}, we present the key results of the \hi mass constraints and the magnification estimates and apply these to speculate on the impact of lensing on future observations. Section~\ref{conclusion} summarises the paper and presents our plan for future work in the field. We assume a Planck 2015 cosmology \citep{planck_2015} throughout. See Table~\ref{tab:summary} for summary of observation details, the source parameters and the our final results. 
\section{OBSERVATIONS AND REDUCTION}\label{sec:obs}
\subsection{Targets} 
All three sources are galaxy-galaxy lenses selected from the Sloan Lens ACS Survey~(SLACS) lens catalog \citep{bolton_2008, auger_2009, newton_2011}. We emphasise that we are interested in detecting \hi in the lensed galaxy and not the lensing galaxy. These strong lenses were identified via a search through the SDSS spectroscopic data for an absorption-dominated spectrum consistent with an early-type galaxy at one redshift and nebular emission lines (Balmer series, {\sc [O\,ii]} and {\sc [O\,iii]}) consistent with a star-forming galaxy at a higher redshift. Given the 3~arcsec fibre diameter, such composite spectra would provide strong evidence of two separate galaxies and hence a lens candidate.
The resulting candidates were observed with the {\it Hubble Space Telescope}~({\it HST}) in bands $V_{\rm HST}$, $I_{\rm 814}$ and $H_{\rm 160}$ to confirm and model the lens. The original SLACS sample consisted of 85 ``grade-A" (i.e. showing clear signs of multiple imaging) lensed systems \citep{auger_2009}, 46 of these were further studied by \citet{newton_2011} to extract information about the source characteristics. These sources lie between $0.2 \le z_{\rm src} \le 1.3 $ with a median modeled (unlensed) stellar mass of $\log_{\rm 10}(M_\star/M_\odot)=8.85$ and a median optical magnification of $8.8$.

\begin{figure}
\begin{center}
\includegraphics[width=1.00\columnwidth]{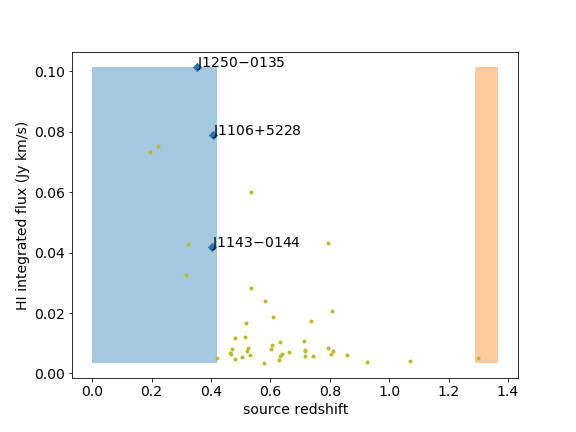}
\caption{Predicted integrated \hi flux as a function of source redshift for SLACS sources described in \citet{newton_2011}, using the $M_\star - M_{\rm HI}$ conversion given in \citet{maddox_2015} and the naive assumption that optical magnification is equal to \hi magnification. Observed sources reported in this work are indicated by the blue diamonds and text, while the other sources in the catalog are shown as yellow dots. The GMRT L-band is shown in blue (identical to the VLA L-band) and the GMRT UHF band is shown in red. \label{fig:slacs_summary}}
\end{center}
\end{figure}

There were several reasons for using the SLACS catalog for \hi candidate selection. First, it is the largest spectroscopic catalog of low to intermediate redshift ($0.2 \lesssim z_{\rm src} \lesssim 1$) lensed galaxies  available. Second, as the targets have strong nebular emission lines, they are star forming and hence may likely have significant cold gas reservoirs. Third, as the lenses had been well modeled in the optical with {\it HST} data, these models could then be utilised for \hi analysis. All three targets in our observations were categorised as ``grade A" lenses  in the catalog.

Fig.~\ref{fig:slacs_summary} shows the estimated \hi integrated flux as a function of redshift, using the  $M_\star - M_{\rm HI}$ correlation\footnote{Specifically, we used a linear interpolation for the $M_\star - M_{\rm HI}$ correlation presented in table~1 in \citet{maddox_2015} for the case which excluded galaxies without SDSS spectra.}\footnote{ Although, the stellar mass function changes significantly from $z\sim0$ to $z\sim0.4$\citep{hopkins_2006}, the \hi mass function is predicted to stay the same \citep{obreschkow_2009_d,lagos_2011}. Therefore, the general trend of the redshift evolution should be towards larger gas reservoirs at a given stellar mass.} presented in \citet{maddox_2015} and the naive assumption that \hi magnification is equivalent to the optical magnification.  The validity of the assumption on the magnification will be explored later in the paper. The predicted \hi masses of our targets (see Table~\ref{tab:summary}) lie in the range $\log_{10}(M_{\rm HI}/M_\odot)=9.5\pm0.5$. 

We observed three SLACS sources with the GMRT: J1106$+$5228, J1250$-$0135 and J1143$-$0144, shown in Fig.~1. Our sources were selected out of the SLACS sample by (1) observability from the GMRT near Pune, India and (2) the predicted magnified \hi integrated flux. The first section of Table~\ref{tab:summary} shows a summary of the observational parameters, including source and lens redshifts.

\subsection{Data reduction} 
In order to benefit from new interferometric data reduction software development, we designed a data processing pipeline using the {\sc stimela}\footnote{https://github.com/SpheMakh/Stimela} interface \citep{makhathini_2018}. This provided a consistent interface over a variety of software tools e.g. {\sc aoflagger} \citep{offringa_2010} , {\sc wsclean} \citep{offringa_2014}, {\sc pybdsf} \citep{mohan_2015} , {\sc casa} \citep{mcmullin_2007}, {\sc cubical} \citep{kenyon_2018} and {\sc meqtrees} \citep{noordam_2010}.

The entire reduction can be roughly divided into three broad sections (known as generations of calibration). In the first, the calibrator fields provide the initial antenna-based complex gain solutions as a function of time and frequency. The frequency-dependent solutions are fixed in at this point. In the second (self-calibration), the target field itself is used to further calibrate the antenna gains, with iterations over decreasing time intervals. Note that the continuum fields contain many bright sources not associated with the target galaxy. It is these objects upon which the self-calibration operates and not the target or lens galaxy which are typically faint.  In the third, we try to minimise artifacts caused by direction-dependent gains imparted to bright off-axis sources \citep{smirnov_2011}. These direction-dependent gain errors result from fluctuations in the primary beam pattern due to antenna pointing errors. The time-dependent gain solutions are only fixed in at the endpoint of the calibration.

The typical duty cycle was 3 minutes on the gain calibrator followed by 15 minutes on the source, with 15 minute scans on the flux (primary) calibrator at the beginning and end of each observation.

The reduction of the calibrator fields made use of the standard {\sc casa}  calibration routines and {\sc wsclean}  for imaging. Both manual and automated flagging was carried out, the latter with {\sc aoflagger} following several tests to find an optimal {\sc aoflagger} strategy. We minimise the impact of radio frequency interference (RFI) on calibration solutions by performing two calibration steps on the calibrators. In the first step, we flag and calibrate. In the second step, we subtract the calibrator source model from the visibilities, flag the residuals and solve again for the calibration solutions. The bandpass solutions showed  fluctuation across the band, especially for the 32~MHz band due to the decrease in channel resolution.

During self-calibration, we used on the order of 5 calibration-source modelling loops. For accurate deconvolution and continuum subtraction in the target fields, we experimented with different combinations of {\sc wsclean} auto-masking and multi-scale settings as well as {\sc pybdsf} for modeling sources as parametric components. RFI flagging with {\sc aoflagger} was used for the target fields as well, but with higher flagging thresholds in order to retain as much data as possible. When we solve for the antenna gain solutions during calibration, we first apply the older (longer time interval) solutions. This ensures that the new gains start closer to an optimised solution, especially as the SNR drops for smaller solution intervals. The SNR of the continuum fields were high enough for this step to be successful and in general the self-calibration step increased the image SNR by approximately a factor of 2. After the final calibration step, the model of the continuum sky is subtracted from the data. 

We now present a brief discussion on the reduction of each individual field. A summary of each field can be found in Table~\ref{tab:summary}.

\clearpage
\begin{table}
\begin{minipage}{\textwidth}
\centering 
\caption{Summary of source properties and derived results for three strongly lensed sources. The {\bf first section} of the table is largely taken from \citet{newton_2011} and is based on SDSS spectroscopy and {\it HST} imaging, although see section~\ref{sec:redshift} for discussion of the source redshifts. The \hi mass predictions are interpolated from \citet{maddox_2015} as described in the text. The impact factor ranges used in the simulations are estimated in section~\ref{sec:parm_sampling}. The {\bf second section} describes the basic observational setup at the GMRT, and the {\bf third section} presents the observational results. The {\bf fourth section} details the derived \hi mass and magnification and constraints (see Fig.~\ref{upperlimits} for a visual representation). The final line presents a \hi mass upper limit using the alternative method discussed in section~\ref{mass_constraints}.}
\begin{tabular}[]{l|lll}
&J1106+5228                             & J1250$-$0135             & J1143$-$0144      \\
\hline
Co-ordinates&11h06m46.15s  +52d28m37.8s&12h50m50.52s  $-$01d35m31.7s&11h43m29.64s  $-$01d44m30.0s\\
$z_{\rm lens}$                         & 0.095                  & 0.087                  & 0.106                  \\
$z_{\rm src}$ (optical)        & $0.4070 \pm 0.0003 $&$ 0.3526 \pm 0.0004$                 & $0.4019 \pm 0.0004$   \\
$z_{\rm src}$ (\hi)            & 0.4073     & 0.3526                              & 0.4019   \\
$\mu$ (optical)                        & 28                     & 13.3                   & 10.4                   \\
$\log_{\rm 10} M_\star/M_\odot$        & $8.72^{+0.23}_{-0.12}$ & $9.37^{+0.14}_{-0.15}$ & $9.00^{+0.23}_{-0.15}$ \\
Einstein radius (arcsec)               & $1.23 \pm 0.14  $ & $1.28 \pm 0.12$            & $1.68 \pm  0.14$      \\
Intrinsic optical isophotal radius (arcsec)      & 0.11                   & 0.25                   & 0.15                   \\
Predicted $\log_{\rm 10} M_{\rm HI}/M_\odot$&$9.4 \pm 0.3$ & $9.7 \pm 0.3$& $9.5 \pm 0.3$\\
Impact factor range (arcsec) &[0, 0.1]& [0, 0.2]&  [0.2, 0.4] \\
\hline
On-source observing time (hours)       & 6.8                    & 16.5                   & 4.5                    \\
Bandwidth (MHz)                        & 4                      & 32                     & 4                      \\
Channel resolution (kHz)               & 8                      & 64                     & 8                      \\
Mean no. of working antennas           & 26                     & 26                     & 28    \\
\hline
RMS of continuum image ($\mu$Jy\,beam$^{-1}$) & 54                     & 16                     & 64      \\
Spectral line \emph{uv}-weighting   & Natural                & Briggs 0.5         & Briggs 0.5         \\
Frequency-integrated flux (JyHz)  &414& 114& -84 \\
RMS of spectral line  (JyHz)   & 110&69&102  \\
\hline
Probability of zero mass ($C_0$)&10$^{-4}$&$0.04$&$0.79$\\
$\langle\mu\rangle$ at optically predicted mass&3.9&2.3&3.3\\
$\langle\log_{\rm 10} M_{\rm HI}/M_\odot\rangle$ &10.2&9.4&-\\
$\langle\mu\rangle (\langle\log_{\rm 10} M_{\rm HI}/M_\odot\rangle)$ &1.9&3.1&-\\
68\% conf. bounds on $\log_{\rm 10} M_{\rm HI}/M_\odot$&[9.5, 10.5]&[8.1, 10.0]&[-, 8.0]\\
95\% conf. bounds on $\log_{\rm 10} M_{\rm HI}/M_\odot$&[7.8, 10.8]&[-, 10.3]&[-, 9.7]\\
99.7\% conf. bounds on $\log_{\rm 10} M_{\rm HI}/M_\odot$&[6.4, 10.8]&[-, 10.4]&[-, 10.1]\\
\hline
$3\sigma$ upper limit $(\log_{\rm 10} M_{\rm HI}/M_\odot)$ &10.7&10.4&10.7\\

\end{tabular}
\label{tab:summary}
\end{minipage}
\end{table}

\clearpage

\subsubsection{J1106+5228 $(z=0.407)$}
This field is dominated by two $\sim0.1$~Jy off-axis point sources, on diametrically opposite parts of the field of view. The two sources are $11$~arcmin and $17$~arcmin away from the phase centre. The half-power radius of the GMRT primary beam at this wavelength is $\sim14$~arcmin. The multiplication (in the image plane) of the time-dependent primary beam pattern with bright, off-axis sources causes errors most prominent in the immediate vicinity of the sources in question. We used a differential gains technique to solve for these sources with  {\sc meqtrees}  and {\sc cubical}, however the limited bandwidth (4~MHz) and source flux meant that the signal to noise ratio (SNR) was too low for differential gains solutions to converge.

\subsubsection{J1250-0135 $(z=0.353)$}
This field is dominated by a bright off-axis radio galaxy with complex, extended morphology not readily seen in previous, lower angular resolution maps. This well-resolved source is situated at an angular distance of $12$~arcmin from the phase centre (i.e. 85 percent of the angular distance to the half-power point). We solved for differential gains solutions towards this problematic source, however the SNR was again too low and source structure too complex to see a significant improvement in the image residuals. The significantly poorer $uv$-coverage for equatorial observations leads to a Point Spread Function (PSF) with significant amplitude outside of the main lobe. The combination of the complex, diffuse source structure, primary beam effects and a PSF with significant side-lobes meant that this source could not be robustly modeled and hence accurately subtracted from the data. For this reason, we chose to extract the spectra of this target with a Briggs 0.5 weighting, suppressing the PSF sidelobes at the expense of a slight sensitivity penalty.

\subsubsection{J1143-0144 $(z=0.402)$}
This is also an equatorial field, which meant that the PSF had high interferometric side-lobes. Due to the presence of large-scale diffuse emission close to the target (likely associated with the parent cluster of the foreground galaxy lens), we choose Briggs 0.5 weighting for the spectral line science with this target, in this case to suppress the shortest spacings and thus the sensitivity to the diffuse foreground emission.

\subsection{Spectral line width and centre}\label{sec:redshift}
The analysis of marginal or non-detections of \hi is complicated by potential misalignment of the \hi and optical emission line centroids in redshift space.  Misalignments can be either due to the measurement uncertainty associated with the spectral lines or intrinsic physical offsets between the emitting components. Intrinsic offsets between the optical and \hi redshifts could occur for a variety of reasons including stellar outflows or asymmetries in the stellar or \hi discs \citep{maddox_2013}. 

There are two additional complications to centroid alignment relevant to this galaxy sample. As the SDSS fibre radius ($\approx 1.5$~arcsec) is comparable to the size of the Einstein radii (see Table~\ref{tab:summary}), some of the emission may not be captured. Furthermore, although lensing is achromatic, intrinsic spatial variation between ionised and neutral gas would result in differing magnifications. An example of a differentially magnified spectrum is shown in Figure 1 of \citet{deane_2015}.

As the original SLACS papers did not quote an uncertainty on the redshift, we re-derive the redshifts along with uncertainties. To do this, we subtract the SDSS model of the foreground galaxy spectrum, and fit a Gaussian profile to the high SNR ({\sc [O\,iii] 5008{\AA}}) line. To factor in the possibility of a misalignment between the optical and \hi centroids as discussed above, we use the width of the profile ($\sim 100$~km/s) instead of the uncertainty on the peak position ($\sim 10$~km/s). The expectations of these re-derived redshifts all agree with the \citet{bolton_2008} redshifts within the either choice of uncertainty.

The full line width, defined as 20 per cent of the peak flux $w_{\rm 20}$, can be predicted from the baryonic Tully-Fisher relation \citep{mcgaugh_2000}. We estimate the baryonic mass using the predicted stellar and \hi masses and the ratio of total gas mass to \hi mass of 1.5 (roughly accounting for molecular gas mass for star forming galaxies at this redshift \citep{geach_2011}). For an edge-on disc, this yields approximately $w_{\rm 20} \sim 200$~km/s for all three sources.

To search for possible detections, the radio spectrum is then convolved with a boxcar of size $\sim 200$~km/s (expected size in the rest frame) and we check for significant peaks within a redshift range of $1\sigma$ from the optical peak.

We then choose two peak thresholds at ${\rm SNR}> 6.5\sigma$ (the same as the class 1 sources in the ALFALFA (Arecibo Legacy Fast ALFA) survey \citep{haynes_2011,jones_2018}) and ${\rm SNR}>3\sigma$, the former representing detections and the latter representing follow-up candidates which show evidence of a marginal detection\footnote{Detection thresholds for \hi sources coincident with optical sources should be set lower than the detections thresholds for a blind survey as the search space is vastly reduced and hence one is less likely to find rare noise spikes.}. Only for the source J1106+5228 do we find a significant peak with ${\rm SNR}>3\sigma$, consistent with the {\sc O\,[iii] 5008} redshift within $1\sigma$, offset from the peak by 65 km/s (see Table~\ref{tab:summary} for optical and \hi redshift centroids). 

We use the integrated flux (i.e. the averaged flux over the 200km/s bin multiplied by the frequency width of the bin) at the optical redshift expectation or at the highest ${\rm SNR}>3\sigma$ peaks if any. The integrated flux values, even the non-detections, are necessary for our novel analysis formalism derived section~\ref{subsec:upper_limits}. 

\section{RESULTS AND SOURCE MODELING}\label{sec:results}
\subsection{Interferometric data products}\label{subsec:obs_result}

\subsubsection{Spectra}\label{sec:spectra}
\begin{figure}
\begin{center}
\includegraphics[width=1.00\columnwidth]{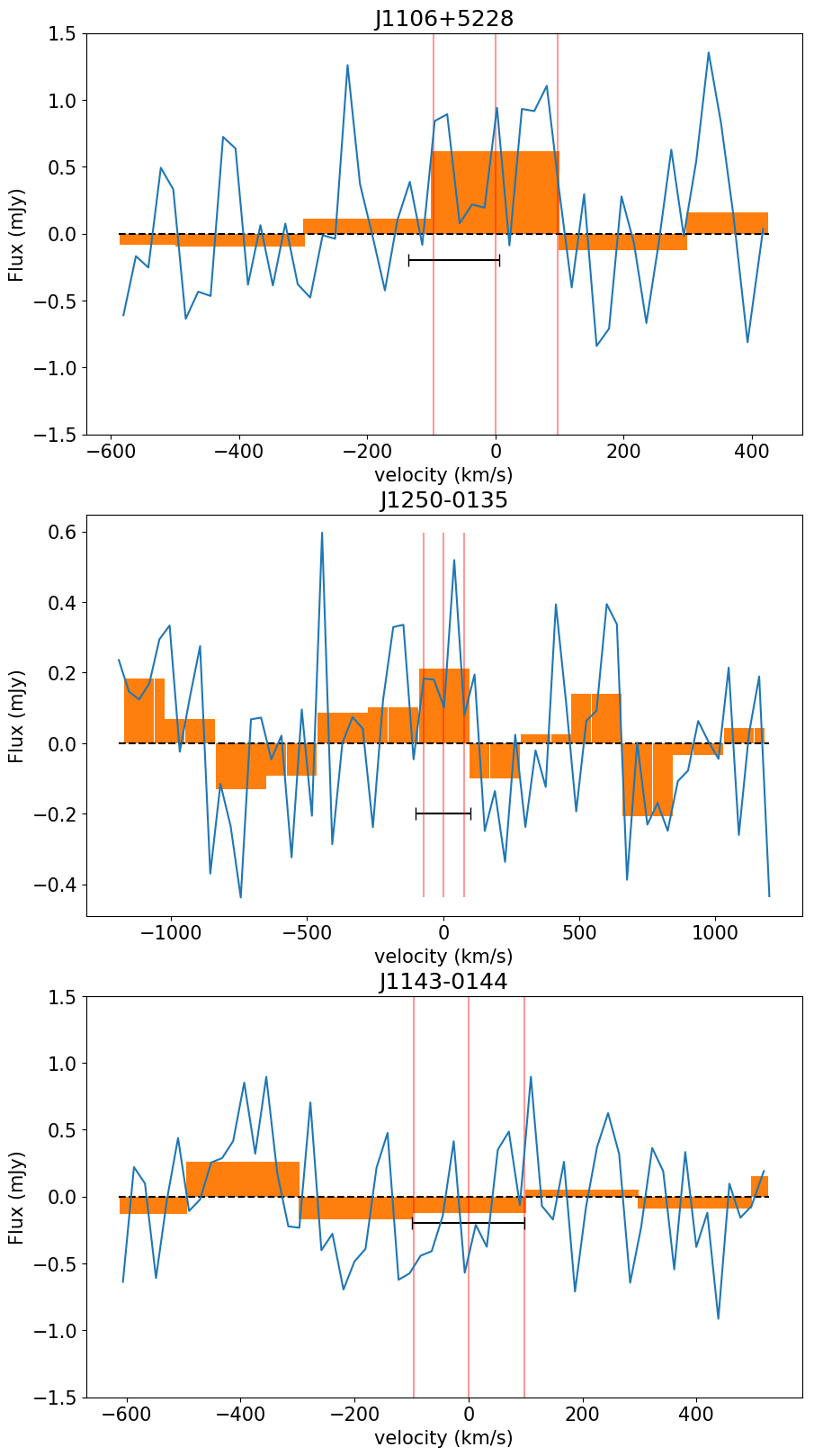}
\caption{Source position-centered spectra with the rest frame line position and 200~km/s line-width indicated by the red vertical lines. The binned spectra are overplotted in orange, where the bin size is set to the number of channels extended by a rest frame velocity of width of 200~km/s at the source redshift. The optical redshift and uncertainty is shown by the horizontal black line. See accompanying text in section~\ref{sec:spectra} and section~\ref{sec:redshift} for more details. Integrated flux and noise estimates are given in Table~\ref{tab:summary}.
\label{binned_spectra}}
\end{center}
\end{figure}

The radio spectra of the targets are shown in Fig.~\ref{binned_spectra}. The binned spectra are overplotted in orange, where the bin size is set to the number of channels extended by a rest frame velocity width of 200~km/s (see section~\ref{sec:redshift} for discussion on the bin width).

An outer \emph{uv}-taper (Gaussian, FWHM of 6~arcsec) was used to maximise sensitivity to extended emission. This angular size was predicted by applying the \hi mass-size relation to the expected \hi mass based on the optical prior. We also subtract the mean of the off-source channels to remove any continuum emission associated with the lens and/or source. 

Following from section~\ref{sec:redshift}, the rest frame line position is set to the expectation of the optical redshift for non-detections, except in the case of J1106+5228 where the spectra has been shifted by 65~km/s to centre the  candidate detection  (integrated SNR of 3.8$\sigma$) at 0~km/s. This offset is within $1\sigma$ of the optical redshift expectation. The optical redshift and uncertainty is shown by the horizontal black line. At this low SNR, we cannot say with high significance that this detection is real, however the evidence available shows that it would be an excellent candidate for follow up observations. Because the integrated SNR is only 3.8, it is statistically impossible to determine further parameters (in addition to the integrated flux), such as the line width (200~km/s) which was set by the optical prior. In contrast to low redshift, high SNR \hi detections, we expect a low SNR detection, rather than attempting to Nyquist sample the putative lensed HI emission line.

The spectral sensitivity and source-centered (frequency and angular position) integrated fluxes are given in Table~\ref{tab:summary}.

\subsubsection{Continuum}
The continuum sensitivities are given in Table~\ref{tab:summary}. The continuum image sensitivity scales approximately as $1/\sqrt{N}$, where $N$ is the number of unflagged visibilities, and is close to the theoretical noise for continuum images indicating that the calibration, flagging and source modelling was successful. Approximately 16 per cent of data was flagged in each hand of polarization due to RFI. Unfortunately, the impact of low-level, broad-band undetected RFI is difficult to estimate and separate from other systematic effects like primary beam errors.

\begin{figure}
\begin{center}
\includegraphics[width=1.00\columnwidth]{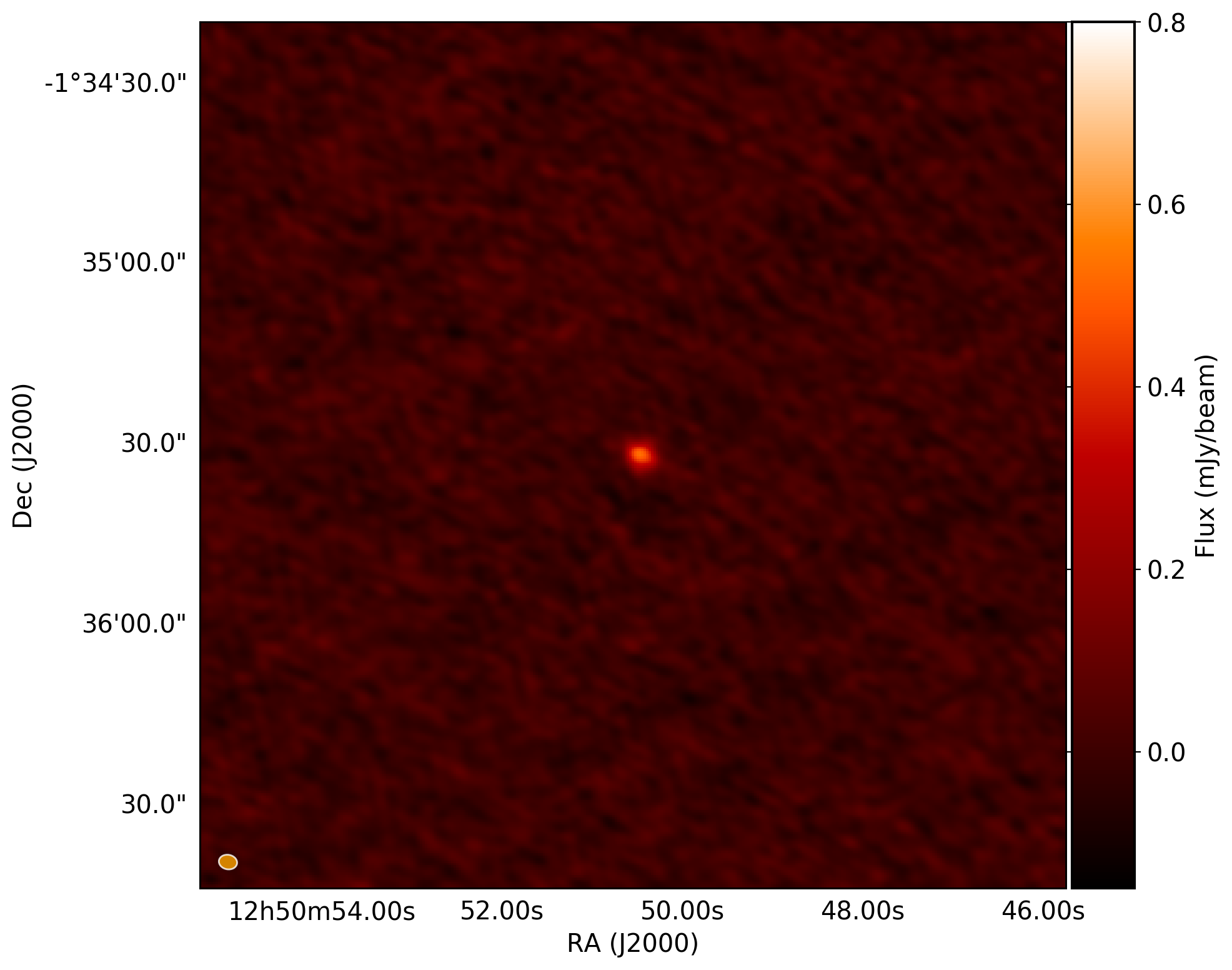}
\caption{Continuum detection coincident with the published {\it HST}-measured position of J1250-0135. The image scale is 2.4$\times$2.4 arcmin$^2$. The source has a peak flux of 0.53~mJy, yielding ${\rm SNR} \sim 30$. The synthesised beam is shown in the bottom left corner. The major and minor axes are 3.0 and 2.4 arcsec respectively.}  
\label{fig:cont_detection}
\end{center}
\end{figure}

We report a continuum detection coincident with the optical position of J1250-0135, shown in Fig.~\ref{fig:cont_detection}. The flux density of this source is $0.53$~mJy and we postulate that the emission originates in the foreground ($z = 0.087$) lens galaxy. The source is unresolved and has a radio luminosity of $L_{\rm 1.4GHz} \sim 10^{23} $~ W/Hz which indicates that the radio component is AGN dominated according to the categorisation of \citet{mauch_2007}. The continuum was not detected for the other two sources.
 
\subsection{HI magnification model}\label{subsec:sims}
We now seek a quantitative prediction of the \hi lensing magnification and hence the expected \hi integrated flux for targets. A theoretical estimate of the \hi magnification requires: (1) a physically-motivated range of possible \hi distributions, (2) an accurate model of the lens and (3) general relativistic ray tracing. We opt for a parametric disc model and explore the dependence of all free parameters on the magnification, marginalising over the nuisance parameters.

\subsubsection{Parametric disc and lens model}
To model the intrinsic \hi surface density $\Sigma_{\rm HI}$,  we adopt the axisymmetric model of \citet{obreschkow_2009b}, 
\begin{equation}
\Sigma_{\rm HI} (r) = \frac{M_{\rm  H}/(2\pi r_{\rm disk}) \exp{(-r/r_{\rm disk})}}{1+R^{\rm c}_{\rm mol}\exp{(-1.6 r/r_{\rm disk})}},
\label{eq:HI_profile}
\end{equation}
where $r$ is the galactocentric radius in the disc plane, $ M_{\rm  H} = M_{\rm H_2}+M_{\rm HI}$,  $r_{\rm disk}$ is the scale length of the neutral hydrogen disk (atomic plus molecular) and $R^{\rm c}_{\rm mol}$ is related to the ratio of molecular to atomic hydrogen mass \citep{obreschkow_2009b} by 
\begin{equation}
 M_{\rm H_2}/M_{\rm HI} = (3.44 R^{\rm c\ -0.506}_{\rm mol}+4.82R^{\rm c\ -1.054}_{\rm mol})^{-1}.
\end{equation}

Interestingly, the HI mass is very tightly correlated to the \hi size  \citep{wang_2016}
\begin{equation}
\log_{\rm 10}(r_{\rm HI}/{\rm kpc})  = 0.506 \log_{\rm 10}(M_{\rm HI}/{\rm M_\odot}) - 3.293,
\label{eq:mass_size}
\end{equation}
where $r_{\rm HI}$ is defined as the radius at which the \hi density drops to $\Sigma_{\rm HI} = 1~{\rm M_\odot pc^{-2}}$. 

We calculate the value of $r_{\rm HI}$ using Equation \ref{eq:mass_size} and then use this to solve for $r_{\rm disk}$ in Equation \ref{eq:HI_profile} for an assumed $M_{\rm HI}$ and $R^{\rm c}_{\rm mol}$. This means that $r_{\rm disk}$ does not need to be sampled separately to $M_{\rm HI}$ and provides important physical consistency.  

For the lens model, we use the single elliptical isothermal sphere model derived in \citet{bolton_2008}. The ray tracing was performed using the {\sc glafic} package \citep{oguri_2010}. As {\sc glafic} only allows for a few input parameteric source types, the \hi discs had to be transformed from being sampled by a grid of pixels to being sampled with a grid of Gaussian distributions. The error $\delta$ introduced by this sampling procedure can be quantified as 
\begin{equation}
\delta \equiv \frac{\int [I_{\rm original} - I_{\rm Gaussian}]}{\int I_{\rm original}},
\end{equation}
where $I_{\rm original}$ and $I_{\rm Gaussian}$ are the intensity distributions of the original and Gaussian sampled images respectively and the integral is over the entire image plane. We find that mean error over all simulations is not significant, $\langle \delta \rangle \approx 10^{-3}$. 
\subsubsection{Parameter sampling and marginalisation}\label{sec:parm_sampling}
The source galaxy in the optical was modeled as a sersic distribution \citep{newton_2011}, however the intrinsic axis ratio, position angle and impact factor  i.e. galaxy centroid with respect to the lens) was not published. To estimate the impact factor for the lensed \hi simulation, we ran a set of Monte-Carlo, lensing simulations. Using {\sc glafic} for ray-tracing and the published optical properties, we find the range of impact factors which can yield the published optical magnification, effectively marginalising over the intrinsic axis ratio and position angle. Given the larger \hi size, this is sufficient for representative \hi simulations as will become clearer later (or see Fig.~\ref{sim_hists}). The ranges of impact factors used are given in Table~\ref{tab:summary}.

To incorporate orientation effects, for each realisation, the two-dimensional disc is simulated, and then randomly rotated in a three-dimensional cube to sample the inclination and position angles of the \hi disc. The position angle is sampled uniformly over the range $[0,2\pi]$. The inclination angle $i$ is sampled with a probability density function (PDF) of $\sin(i)$ over the range $[0,\pi/2]$. 

We sample $\log_{\rm 10}(R^{\rm c}_{\rm mol})$ from a normal distribution with $[{\rm mean, stdev}] = [-0.1,0.3]$. This is consistent with the range of $M_{\rm H_2}/M_{\rm HI}$ quoted in \citet{catinella_2018} for the stellar mass range of these galaxies. 

For the lens, the Einstein radius is sampled from a Gaussian distribution. The ellipticity and position angle of the lens is set to that of the observed optical distribution.
 
\subsubsection{Dependence of free disc parameters on magnification}
In order to understand the dependence of the magnification on the source parameters. We calculate \hi magnifications for $10^4$ random samples of the model and calculate the cumulative probability density function (CDF) of $\mu$ for each parameter individually, marginalised over the remaining parameters.  These marginalized functions are computed numerically using a Monte Carlo integrator, which we use for a non-parameteric estimation of the expectation and confidence intervals, as shown in Fig.~\ref{sim_hists} for J1106+5228 and in Appendix A for the other two sources.

To determine the PDF of $\mu(M_{\rm HI})$, we marginalise over the other free parameters. This is the only function which we need and it requires no PDF for $M_{\rm HI}$ (in other words, the PDF of $\mu$ is calculated at each $M_{\rm HI}$). This is presented in the upper-left panel of Fig.~\ref{sim_hists} and exhibits a tight relation between the average magnification and the \hi mass which arises from Equation~\eqref{eq:mass_size}. As the magnification is almost completely dependent on the mass, we can calculate a simplified conversion between \hi mass and integrated \hi flux.

To calculate the dependence of $\mu$ on the other disc parameters, we marginalise over $M_{\rm HI}$ by sampling from the PDF predicted from the optical prior.

\begin{figure}
\begin{center}
\includegraphics[width=1.00\columnwidth]{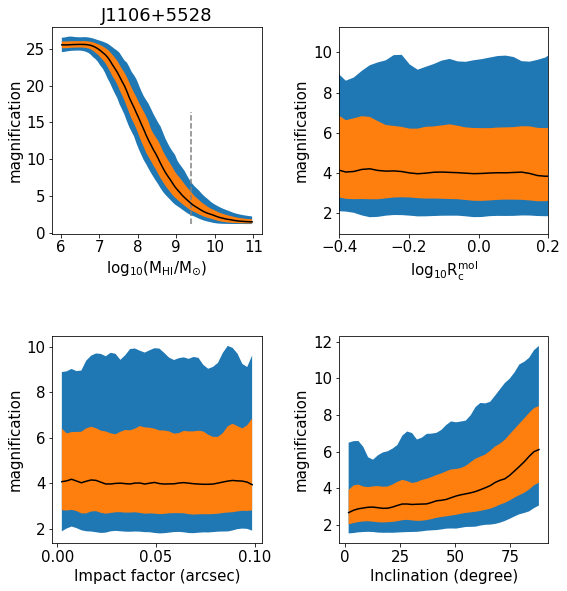}
\caption{A selection of bivariate relations obtained from $10^4$ \hi lensing simulations of the J1106+5228 lensing system. In each panel, the black curve shows the expectation, while the orange and blue filled areas show the 68 and 95 percent confidence intervals respectively. {\bf Upper left:} magnification as a function of $M_{\rm HI}$. The gray, dashed vertical line presents the mass prediction from \citet{maddox_2015} stellar-\hi mass relation. In this panel $M_{\rm HI}$ was sampled from a uniform distribution (i.e. no PDF), however, in the other panels $M_{\rm HI}$ was sampled from the optical prior. The $M_{\rm HI}$-magnification coupling is a direct result of Equation \ref{eq:mass_size}.
{\bf Upper right:} magnification as a function of $R_{\rm c}^{\rm mol}$, showing no correlation is apparent over this range of parameters.
{\bf lower left:} magnification as a function of impact factor for the range of impact factors as estimated from the {\it HST} model, showing this is indeed a nuisance parameter in this particular analysis. 
{\bf lower right:} magnification as a function of inclination, an increase in magnification is evident as the source becomes increasingly inclined. See main text for further detail on Monte Carlo assumptions, particularly on chosen parameter distributions and the justifications thereof.} \label{sim_hists}
\end{center}
\end{figure}

\subsection{Probability distribution of \hi mass}\label{subsec:upper_limits}
We now describe a mathematical model to evaluate the \hi mass probability distribution, given an estimate of the integrated source flux and noise. This is a simplified model in which the measured integrated flux is solely attributed to Gaussian noise and lensed \hi emission.

Under this assumption, the PDF of the real (i.e. noise-free) \hi frequency-integrated flux $S$ is given by
\begin{equation}
\rho(S) = \frac{1}{\sqrt{2\pi\sigma_{S}^2}} \exp\left(-\frac{(S-S_0)^2}{2\sigma_{S}^2}\right),
\label{eq:rho_S}
\end{equation}
where $S_0$ is the measured source frequency-integrated flux and $\sigma_{S}$ is the standard deviation of the frequency-integrated flux. The units of $S$ are JyHz=$10^{-26}$~W/m$^2$ and we measure $S_0$ and $\sigma_S$ from the spectra (see Table~\ref{tab:summary} for the measured values, Fig.~\ref{binned_spectra} for the spectra and its accompanying text for details of the measurement).

We do {\bf not} assume that a detection has been made. Instead we attempt to answer the question: What is the probability distribution of real integrated flux (i.e. that is not due to noise) and associated mass? The equations are generally true for a measurement of a quantity $S$, given an integrated signal $S_0$ with Gaussian noise. Importantly, we define ``signal" as the integrated flux, irrespective of a ``detection", hence the signal can be lower than the noise and even negative. In the case of pure noise, the expectation of the measured integrated flux is $<S_0>\ =0$. If there is a true signal (even if it is smaller than the noise $\sigma_S$), $<S_0>$ becomes positive. However, if $S_0/\sigma_S$ is small (as in two of our galaxies), the PDF of Equation~\eqref{eq:rho_S} will still have its expectation at $S_0$, but with a large uncertainty that accounts for a significant probability of there being no true signal.

Using the parametric lensed \hi model described in section~\ref{subsec:sims}, the frequency-integrated flux $S$ at a given mass (marginalised over the other free parameters) is equal to (see \citet{meyer_2017} for the expression for an unlensed galaxy)
\begin{equation}
 S  = \frac{\langle\mu\rangle(M_{\rm HI}) M_{\rm HI}}{49.7 D_{\rm L}^2},
\label{eq:S}
\end{equation}
where the expectation of the magnification at a given mass $\langle\mu\rangle(M_{\rm HI})$ is computed numerically from the Monte Carlo simulations and  $M_{\rm HI}$ is in units of $M_\odot$. $D_{\rm L}$ is the luminosity distance in units of Mpc and is calculated from the optical spectroscopic redshift of the lensed source. 

By accounting for the $M_{\rm HI}$-dependent lensing factor using our simulation results, we convert $\rho(S)$ into a PDF for the \hi mass

\begin{equation}
\rho(M_{\rm HI}) = \rho(S) \dfrac{{\rm d} S}{{\rm d} M_{\rm HI}}.
\label{eq:rho_mhi}
\end{equation}
Differentiating Equation~\eqref{eq:S} with respect to the \hi mass
\begin{equation}
\dfrac{{\rm d} S}{{\rm d} M_{\rm HI}}  = \frac{1}{49.7 D_{\rm L}^2} \left(\langle \mu \rangle(M_{\rm HI}) + M_{\rm HI} \dfrac{{\rm d} \langle\mu\rangle(M_{\rm HI})}{{\rm d} M_{\rm HI}} \right),
\end{equation}
where ${\rm d} \langle\mu\rangle(M_{\rm HI})/{\rm d} M_{\rm HI}$ can be calculated numerically. 

In the presence of noise, common objects (e.g. low mass \hi galaxies) can be mistaken as rare objects (e.g. high mass \hi galaxies) and vice-versa. However, as there are more common objects, the number of common objects being mistaken for rare objects is larger than the reverse case. This leads to an over-estimation of the number of rare objects. This is known as the Eddington bias \citep{eddington_1913}. Given the asymmetric, declining \hi mass function (HIMF) and symmetric Gaussian noise, it is much more likely that a source of measured integrated \hi flux $S_0$ has a true integrated flux smaller than $S_0$ and a positive measurement error  than vice versa. 

Another way of understanding this is to consider a random sample $x=\log(M)$ drawn from a steep (i.e. highly asymmetric) mass function $\rho(x)$. Then perturb $x$ by a random error drawn from a Gaussian. Over many samples, these random errors systematically move the mass function towards the high-mass end. This is an important systematic effect, whenever the mass function changes significantly over the uncertainty of the mass measurement.

We correct for this bias, by using the HIMF as a population prior following \citet{obreschkow_2018}. The de-biased mass PDF is given by:
\begin{equation}
\rho_{\rm Edd}(M_{\rm HI}) = \frac{\rho(M_{\rm HI})\phi(M_{\rm HI})}{\int_0^\infty \rho(M'_{\rm HI})\phi(M'_{\rm HI}){\rm d}M'_{\rm HI}},
\end{equation}
where $\phi(M_{\rm HI})$ is the HIMF, expressed in linear rather than logarithmic units, i.e.~$\phi(M_{\rm HI})={\rm d}n(M_{\rm HI})/{\rm d}M_{\rm HI}$. We use the ALFALFA-100\% HIMF \citep{jones_2018}. We note that this HIMF was measured at $z\approx0$, and that the HIMF at $z=\sim0.4$ may be slightly different \citep[e.g.][]{obreschkow_2009c,lagos_2011}.

We estimate confidence intervals of the \hi mass by forming the cumulative probability density 

\begin{equation}
\Phi_{\rm Edd}(M_{\rm HI}) = C_0 +  \int_0^{M_{\rm HI}} \rho_{\rm Edd}(M_{\rm HI}^\prime){\rm d}M_{\rm HI}^\prime,
\end{equation}
where $C_0$ is the probability that the real integrated flux $S$ is negative, 
\begin{equation}
C_0 = \int_{-\infty}^0\rho(S){\rm d}S~=\frac{1}{2}\left[1+{\rm erf}\left(\frac{-S_0}{\sigma_s\sqrt{2}}\right)\right],
\end{equation}

As physical considerations do not permit $M_{\rm HI}$ to be negative, $C_0$ should be interpreted as the probability of zero integrated \hi flux.

\begin{figure}
\begin{center}
\includegraphics[width=1.00\columnwidth]{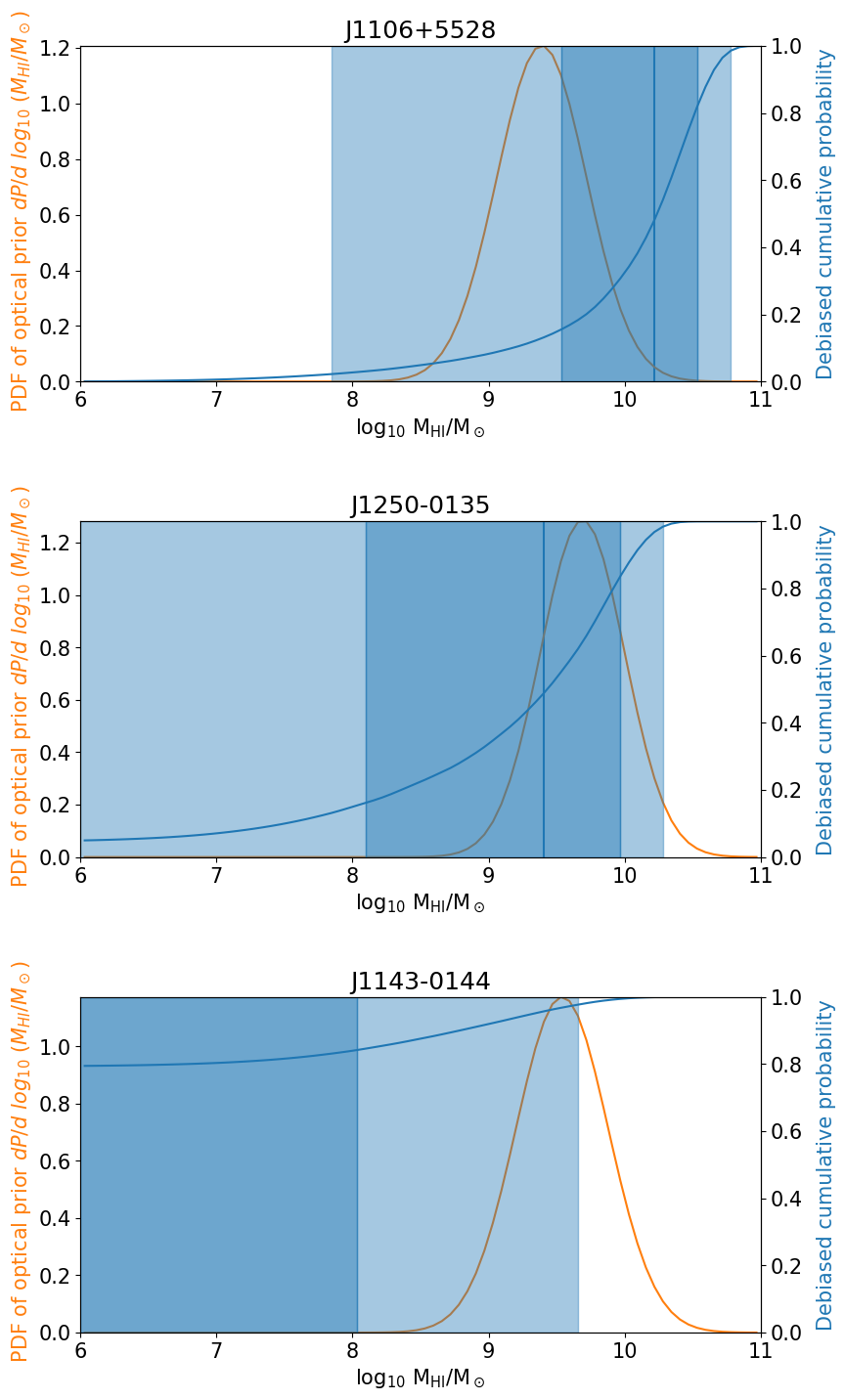}
\caption{\hi mass probability density functions for the three observed sources. The probability density as predicted by the \citet{maddox_2015} stellar mass\,- \hi mass correlation is shown with orange curve. The Eddington-bias corrected (HIMF prior), cumulative  probability $\Phi_{\rm Edd}(M_{\rm HI})$ is shown by the blue curve. The 68 per cent and 95 per cent confidence intervals of $\Phi_{\rm Edd}(M_{\rm HI})$ are shown with dark and light blue shading respectively. The vertical blue line shows the expectation of the \hi mass. \label{upperlimits}}
\end{center}
\end{figure}

The de-biased cumulative probability $\Phi_{\rm Edd}(M_{\rm HI})$ (along with 68 percent and 95 percent confidence intervals) and the PDF of the \hi mass derived from the stellar mass is shown in Fig~\ref{upperlimits}. Where applicable, the expectation of the de-biased mass as well as the boundaries of the confidence intervals are given in Table~\ref{tab:summary}.

\section{DISCUSSION}\label{discussion} 
\subsection{\hi mass constraints}\label{mass_constraints}
Our Bayesian formalism aims to extract the maximum amount of information possible from the radio measurement. This is achieved by leveraging the optical spectroscopic information, correcting for Eddington bias and folding in a physical model of the magnification. The final product is the cumulative mass probability distribution along with non-parametric uncertainties. We emphasise that, these results apply only to the sampled 200~km/s (rest frame) bin centered on the best estimate of the source frequency position (see section~\ref{sec:redshift}) and relate to the total \hi mass of the galaxy as far as this bin contains the majority of the \hi flux. In the event of unsampled flux, these results would underestimate the total \hi mass and, on average, shift the cumulative probability  $\Phi_{\rm Edd}(M_{\rm HI})$ to lower masses.

We report evidence suggesting a marginal $(3.8\sigma)$ \hi detection for the lensed galaxy J1106+5228 (z=0.4073) offset by 65~km/s from the optical redshift. We estimate the intrinsic mass integrated over 200km/s (rest-frame) at $\log_{\rm 10} (M_{\rm HI}/M_\odot) = 10.2^{+0.3}_{-0.7}$ within the 68 per cent confidence interval. This estimated mass range is consistent with the stellar-mass prediction (see Fig.~\ref{upperlimits}).

For all three sources, we do not find unambiguous detections at the expectation of the optical redshift but still extract information on $M_{\rm HI}$. For J1250$-$0135, we estimate $\log_{\rm 10} (M_{\rm HI}/M_\odot) = 9.4^{+0.6}_{-1.3}$ within a 68 per cent confidence interval. This is consistent with the optical prediction. For J1143$-$0144 we obtain a $2\sigma$ upper limit of $\log_{\rm 10} (M_{\rm HI}/M_\odot) = 9.7$ (see Table~\ref{tab:summary} for the full results).

In this paper, we have estimated that the source can be well-sampled by a Gaussian of ${\rm FWHM} \sim 6$~arcsec over a 200~km/s frequency interval. However, the physical size of the \hi disc is dependent on the \hi mass \citep{wang_2016} and the frequency range is dependent on the inclination and total galaxy mass \citep{mcgaugh_2000}. Future analyses could be improved by a more detailed sampling of the integrated flux as a function of intrinsic mass and inclination. 

The only previously published upper-limit on integrated lensed \hi flux \citep{hunt_2016} placed competitive $3\sigma$ $M_{\rm HI}$ upper limits of $ 6.58 \times 10^9 M_\odot$ at $z = 0.398$ and $1.5 \times 10^{10}  M_\odot$ at $z = 0.487$, however the assumption was made that the \hi magnification is equal to the optical magnification which is inconsistent with our simulations (see next section). Their method constrains the \hi mass by taking a factor of the spectral RMS as an upper limit on the lensed \hi signal. We compare the $3\sigma$ (99.7\% confidence) limits derived by the two methods in the bottom two rows of Table~\ref{tab:summary}. One difference is that in our model, the upper limits are a monotonically increasing function of the source-centered integrated flux. This implies that even negative integrated flux contains information about the possible source mass by lowering the \hi mass upper limits as in the case of J1143-0144.

\subsection{The HI magnification factor}
For all three systems, the magnification is strongly dependent on the \hi mass and the relation follows a reversed-`S' shape curve (see Fig.~\ref{sim_hists} upper left panel). On the low-mass end, the magnification converges to that of a point source (similar to the optical magnification) and on the high-mass end the magnification converges to 1. Between these extremes, the \hi magnification is a monotonically decreasing function of \hi mass. This is because the \hi mass-size relation is monotonically increasing and the magnification is approximately equal to the ratio of lensed-to-intrinsic angular size.

Comparing the \hi magnifications at the optically predicted \hi mass with the optical magnifications (see Table. ~\ref{tab:summary}), we see that the \hi magnifications are predicted to be significantly lower than the optical magnifications by a factor $\sim 3-7$. The general trend of lower \hi magnifications is due to \hi being more extended than the stellar component. In practice, this effect should be heightened for these optically-selected lensed systems which are biased towards compact nebular line emission components. Moreover, the effect is enhanced at these intermediate redshifts given the large \hi source size-to-Einstein radius ratio.

As \hi is more extended, the dependence of the magnification on sub-arcsecond offsets of the source centroid is significantly reduced. This point is illustrated in Fig.~\ref{sim_hists1143} (upper left panel), where there is only significant fluctuation in the magnification due to changes in the impact factor at very small masses (i.e. small sizes). This implies that the \hi magnifications should not be approximated by the optical magnification but must be modeled separately. Note that this is not seen for the other two objects as the impact factor ranges were estimated to be closer to zero. 
\subsection{Considerations for future observations and surveys}
Our results suggest that at a redshift of $z\sim0.4$, there is potential for using targeted observations of strong lenses with Einstein radii on the order of $\sim 1-2$~arcsec to push the highest-redshift \hi detection threshold, as J1106+5228 would be if confirmed. However, due to the decreasing magnification boost as mass increases (see Fig.~\ref{sim_hists}), we exclude the scenario of a large mass coupled with a high magnification. Again, these statements are only valid for SLACS-selected lenses at these intermediate redshifts for $\sim1$~arcsec-scale Einstein radii.

Strong lenses will however become increasingly important to consider for future \hi surveys. There are several factors to consider. Firstly, as the source redshift increases the lensing optical depth increases (i.e. more sources are lensed). For a velocity-integrated \hi flux cut of $1.0$~mJy~km/s, the fraction of lensed galaxies out of all galaxies will increase by a roughly 2-3 orders of magnitude from $z\sim 0.4$ to $z\sim 2$ \citep{deane_2015}. Secondly, the high end of the \hi mass function might move to smaller masses with increasing redshift \citep{lagos_2011} which by Equation \ref{eq:mass_size} would mean smaller intrinsic sizes and hence higher magnifications. Thirdly, the angular scale increases from approximately 5~kpc/arcsec to approximately 8 kpc/arcsec, which is an effective decrease of about 2.5 in solid angle of the source, increasing magnification significantly for all source masses.

\section{CONCLUSIONS AND FUTURE WORK}\label{conclusion} 
This work presents the first targeted interferometric observations of strongly lensed \hi in emission, as well as the first detailed predictions of integrated \hi flux magnification in individual galaxy-galaxy lensing systems. We have also developed a Bayesian formalism to estimate the \hi mass probability density functions for all sources, even the clear non-detections. The spectrum of source J1106+5228 shows evidence of a marginal detection and is therefore an excellent candidate for follow up observations.
 
In the theory component of this work, we show that for this class of lensing system, the \hi magnification is a monotonically decreasing function of \hi mass because the \hi mass-size relation is monotonically increasing. There is also saturation at low mass as the disc approximates a point source and at high mass where the disc is much larger than the Einstein radius.

The \hi lensing simulation toolkit presented here allows for realistic feasibility studies for planning observations of \hi in galaxy-galaxy lenses. We continue this lensed-\hi campaign with both the upgraded-GMRT and the MeerKAT telescopes \citep{deane_2016}.

In future, we look to extend the analysis to include cluster lenses as well as the statistics of lensing in cosmological volumes which would predict the effect of \hi lensing on next-generation SKA surveys and the observed \hi mass function, with particular reference to blind \hi lens selection.

\clearpage
\appendix
\section{Extended simulation results}
We present the simulation results for sources J1250-0135 and J1143-0144 as described in section~\ref{subsec:sims}. While the current observations of these targets do not share the heightened interest of a possible marginal detection as is the case with J1106+5228, the trends illustrate some of the relevant caveats to be considered in \hi lensing, particularly for these low-to-intermediate redshifts.
\begin{figure}
\begin{center}
\includegraphics[width=1.00\columnwidth]{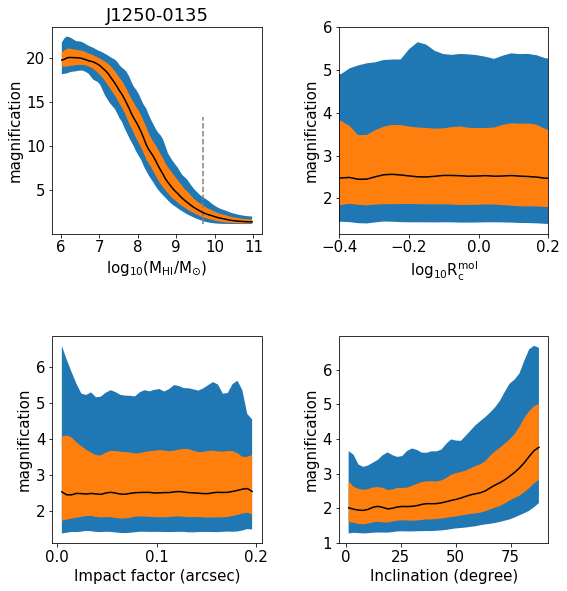}
\caption{A selection of bivariate relations obtained from $10^4$ \hi lensing simulations for the J1250$-$0135 galaxy-galaxy lens. In each panel, the black curve shows the expectation, while the orange and blue filled areas show the 68 and 95 percent confidence intervals respectively.  {\bf Upper left:} magnification as a function of $M_{\rm HI}$. The gray, dashed vertical line presents the mass prediction from Maddox stellar-\hi mass relation. In this panel $M_{\rm HI}$ was sampled from a uniform distribution, however, in the other panels $M_{\rm HI}$ was sampled from the predicted mass probability density.
{\bf Upper right:} magnification as a function of $R_{\rm c}^{\rm mol}$, showing no correlation is apparent over this range of parameters.
{\bf lower left:} magnification as a function of impact factor for the range of impact factors as estimated from the {\it HST} lens model. 
{\bf lower right:} magnification as a function of inclination. See main text for further detail on Monte Carlo assumptions, particularly on chosen parameter distributions and the justifications thereof.
\label{sim_hists1250} }
\end{center}
\end{figure}

\begin{figure}
\begin{center}
\includegraphics[width=1.00\columnwidth]{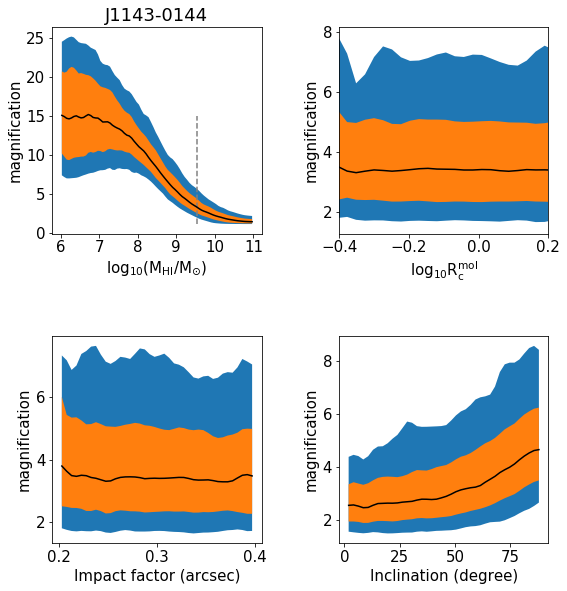}
\caption{A selection of bivariate relations obtained from \hi lensing simulations for the J1143$-$0144 galaxy-galaxy lens. See the caption of Fig.~\ref{sim_hists1250} for more details.
\label{sim_hists1143} }
\end{center}
\end{figure}
\clearpage
\section*{Acknowledgements}
We thank Natasha Maddox and Julia Healy for discussions on alignment of the optical and \hi signals in the frequency domain. We thank Gyula Jozsa for conversations on \hi science. We also thank Benjamin Hugo, Kshitij Thorat, Sphesihle Makhathini, Jonathan S. Kenyon and Oleg Smirnov for advice on interferometric data reduction. We thank Andre Offringa for help with using the {\sc wsclean} software. This research was supported by the South African Radio Astronomy Observatory, which is a facility of the National Research Foundation, an agency of the Department of Science and Technology.

\bibliography{biblio.bib}

\end{document}